\documentclass[12pt]{article}

\usepackage{graphicx}
\usepackage{longtable}

\begin{document}
\title{On the Consistency of the Ladder Approximation and the
Rainbow Approximation of Dyson-Schwinger Equation of QCD}

\author{{Lei Chang$^{1}$, and Yu-xin Liu$^{1,2,3,}$\thanks{Corresponding Author} }\\[3mm]
\normalsize{$^1$ Department of Physics and the MOEC Key Laboratory
of Heavy Ion Physics, } \\
\normalsize{Peking University, Beijing 100871, China}\\
\normalsize{$^2$ CCAST (World Laboratory), P.O. Box 8730, Beijing
100080, China} \\
\normalsize{$^3$ Center of Theoretical Nuclear Physics, National
Laboratory of}\\ \normalsize{ Heavy Ion Accelerator, Lanzhou
730000, China}  }

\maketitle



\begin{abstract}
We study the consistency of the ladder approximation and the rainbow
approximation of the Dyson-Schwinger equation of QCD. By considering
the non-Abelian property of QCD, we show that the QED-type
Ward-Takahashi identity is not required for the rainbow-ladder
approximation of QCD. It indicates that there does not exists any
internal inconsistency in the usual rainbow-ladder approximation of
QCD. In addition, we propose an modified ladder approximation which
guarantees the Slavnov-Taylor identity for the quark-gluon vertex
omitting the ghost effect in the approximation.

\end{abstract}

{\bf PACS numbers:} 11.15.Tk, 12.38.Lg, 12.38.Aw



\newpage

\parindent=20pt


Studies of the truncated sets of Dyson-Schwinger equations have long
been pursued as a possible basis for studying nonperturbative
aspects of quantum field theories\cite{Pagels}. The simplest
truncation scheme is the so-called rainbow-ladder approximation,
which has been used extensively and successfully to study various
nonperturbative aspects of strong interaction physics (see for
example
Refs.~\cite{Craig01,Alkofer01,Craig02,Craig03,Fischer06,Watson01,Craig04,Alkofer02,Craig05,Liu0135,Zong03,Pennington04,Alkofer04,YCL06}).
Recently, the author of Ref.~\cite{Gogohia} claims that the ladder
approximation to QCD is internally inconsistent with a QED-type
Ward-Takahashi relation, then all the results based on the
nontrivial solutions of the quark Dyson-Schwinger equation in the
ladder approximation should be reconsidered, and its use in the
whole energy/momentum range should be abandoned. Since the
rainbow-ladder approximation is such a widely used approximation,
its internal consistency is then an imperative issue and deserves
careful investigation. Then this claim is argued by the authors of
Ref.~\cite{zong}, where they pointed out that the Ward-Takahashi
relation could not be required in the ladder approximation to QCD
and the claim in Ref.~\cite{Gogohia} is not correct at least in the
case of making use of the weak coupling expansion. It is clear that
there exists a contradiction between the Ward-Takahashi identity
omitting the ghost effect and the ladder approximation in view of
the arguments in Ref.~\cite{zong}. However, an explicit and general
verification for the contradiction to be allowed has not yet been
given. There leaves then still a room to prove and generalize the
reason mentioned in Ref.~\cite{zong}. One of the aims of this letter
is just to give an exact expression for the contradiction and to
show that the contradiction is allowed physically to the usual
ladder approximation. Then we would propose a modified ladder
approximation to conciliate the contradiction. In the following we
shall first briefly recall the arguments in Refs.~\cite{Gogohia} and
\cite{zong}, and then give our statements for the problem.

Let us begin with the rainbow approximation of the quark equation
and the ladder truncation of the quark-gluon vertex equation. For a
given quark flavor, the quark Dyson-Schwinger equation (DSE) under
the rainbow approximation in QCD is expressed \cite{fnEuclidean}


\begin{equation}\label{raindse}
S(p)^{-1} =  i\gamma\cdot p + m_{0} + g^{2}C_{F}\! \int \!
\frac{d^{4}q}{(2\pi)^{4}}D_{\mu\nu}(p\! - \! q)
\gamma_{\mu}S(q)\gamma_{\nu}\, ,
\end{equation}
where $m_{0}$ is the current quark mass, $C_{F}$ is
the eigenvalue of the quadratic Casimir operator in the fundamental
representation and $D_{\mu\nu}(k)$ is the dressed-gluon propagator.
The DSE of the quark-gluon vertex under the ladder approximation at
zero momentum transfer reads
\begin{equation}\label{ladddse}
\Gamma_{\sigma}^{a}(p) = i\gamma_{\sigma}T^{a} -g^{2} \! \int \!
\frac{d^{4}q}{(2\pi)^{4}} D_{\mu\nu}(p\! - \!
q)T^{b}\gamma_{\mu}S(q)\Gamma_{\sigma}^{a}(q)
S(q)T^{b}\gamma_{\nu}\, , \;
\end{equation}
where $a,b$ are color indices with $T^ {a}$ the standard Gell-Mann
$SU(3)$ representation. It is noted that only the one gluon exchange
between a quark and an antiquark is allowed in such a ladder
approximation, and this is usually referred to as Abelian
approximation. It has been shown that, for color singlet vertex(such
as vector, axial-vector and pseudoscalar), the Abelian ladder
approximation is valid because of some Ward-Takahashi identities.
However, in the case of quark-gluon vertex, the non-Abelian effect
would be included in the expression of its equations. There would
then exist a contradiction between the Abelian ladder approximation
and the Ward-Takahashi identity if one omit the non-Abelian effect.
Another fundamental expression is the differential form of
Eq.(\ref{raindse}), i.e.
\begin{equation}\label{diffdse-1}
\frac{\partial S(p)^{-1}}{\partial p_{\sigma}}=i\gamma_{\sigma}+
g^{2}C_{F} \!\! \int \!\! \frac{d^{4}q}{(2\pi)^{4}}D_{\mu\nu}(p-q)
\gamma_{\mu}\frac{\partial S(q)}{\partial q_{\sigma}}\gamma_{\nu}\,
.
\end{equation}
We also note that this differential form of quark equation is just
the special one in the above mentioned ladder approximation. With
Eqs.(\ref{raindse}), (\ref{ladddse}) and (\ref{diffdse-1}), we can
recall the arguments in Refs.~\cite{Gogohia} and \cite{zong}.

The author of Ref.~\cite{Gogohia} argues that, in the ladder
approximation, one should omit the ghost-quark scattering kernel
contained in the Slavnov-Taylor identity of QCD and therefore the
Slavnov-Taylor identity is reduced to the Ward-Takahashi relation
\begin{equation}\label{WardI}
\Gamma_{\sigma}^{a}(p)=T^{a}\frac{\partial S^{-1}(p)}{\partial
p_{\sigma}} \, ,
\end{equation}
which should be valid in the ladder approximation to QCD. Using
the above Ward-Takahashi identity and comparing
Eqs.(\ref{ladddse}) and (\ref{diffdse-1}), one can obtain a
relation
\begin{equation}\label{relation}
-\frac{1}{2}g^{2}C_{A}\int\frac{d^{4}q}{(2\pi)^{4}}D_{\mu\nu}(p-q)
\gamma_{\mu}\frac{\partial S(q)}{\partial
q_{\sigma}}\gamma_{\nu}=0 \, ,
\end{equation}
where $C_{A}$ is the eigenvalue of the quadratic Casimir operator in
the adjoint representation. From the above relation, the author
obtains,
\begin{equation}
\partial_{\sigma}\Sigma(p)=0 \, ,
\end{equation}
for the quark self-energy
\begin{equation}\label{self-energy}
\Sigma(p)=\int\frac{d^{4}q}{(2\pi)^{4}}D_{\mu\nu}(p-q)
\gamma_{\mu}S(q)\gamma_{\nu} \, ,
\end{equation}
and concludes that, in the ladder approximation of QCD, the quark
propagator is only the free one, apart from a redefinition of quark
mass, i.e. there is no running/dressed quark mass and in turn the
ladder approximation is internally inconsistent.

The authors of Ref.~\cite{zong} note that Eq.~(\ref{relation}) may
not be true in the ladder approximation. Based on the weak coupling
expansion in Feynman gauge, they obtain mathematically an analytic
result, which reads
\begin{equation}\label{zong-result}
-\frac{1}{2}g^{2}C_{A}\int\frac{d^{4}q}{(2\pi)^{4}}D_{\mu\nu}(p-q)
\gamma_{\mu}\frac{\partial S(q)}{\partial q_{\sigma}}
\gamma_{\nu}\neq 0 \, .
\end{equation}
Furthermore they argue that Eq.~(\ref{zong-result}) is true, because
a lot of numerical calculations with suitable model gluon propagator
in specific gauge have given nonzero results. Since the relation in
Eq.~(\ref{relation}) is directly the result of Eq.~(\ref{WardI}),
the Ward-Takahashi identity should then not be required because the
relation in Eq.~(\ref{relation}) is not true in weak coupling
expansion at least. It is evident that such an argument made in
Ref.~(\cite{zong}) is not solid enough due to a lack of general
verification in view of strong interaction physics (QCD). We then
try to prove the validity of Eq.~(\ref{zong-result}) in a general
point of view.

%
Due to the property of non-Abelian gauge theory, the gauge fields of
QCD are the self-interacting ones. The derivative of the gluon
propagator with respect to momentum is related to the three-gluon
vertex based on the Slavnov-Taylor identity omitting the ghost
effects and can be written as~\cite{Bhagwat-Tandy04}
\begin{equation}\label{relation-2}
\Gamma_{\mu\nu\sigma}^{3g}(k,k)=-\frac{\partial}{\partial
k_{\sigma}}D_{\mu\nu}^{-1}(k) \, .
\end{equation}
With such a relation, one has
\begin{eqnarray}
& &
-\frac{1}{2}g^{2}C_{A}\int\frac{d^{4}q}{(2\pi)^{4}}D_{\mu\nu}(p-q)
\gamma_{\mu}\frac{\partial S(q)}{\partial q_{\sigma}}\gamma_{\nu}
\nonumber \\ &  = &
\frac{1}{2}g^{2}C_{A}\int\frac{d^{4}q}{(2\pi)^{4}}\Big\{
\frac{\partial D_{\mu\nu}(p-q)}{\partial q_{\sigma}} \gamma_{\mu}
S(q)\gamma_{\nu} - \frac{\partial}{\partial q_{\sigma}}\big[
D_{\mu\nu}(p-q) \gamma_{\mu} S(q) \gamma_{\nu} \big] \Big\}
\nonumber \\
& = & -\frac{1}{2}g^{2}C_{A}\int\frac{d^{4}q}{(2\pi)^{4}}
\frac{\partial D_{\mu\nu}(k)}{\partial k_{\sigma}} \gamma_{\mu}
S(q)\gamma_{\nu}   \nonumber \\
& = & -\frac{1}{2}g^{2}C_{A}\int\frac{d^{4}q}{(2\pi)^{4}}
D_{\mu\alpha}(k)\Gamma_{\alpha\beta\sigma}^{3g}(k)D_{\beta\nu}(k)
\gamma_{\mu}S(q)\gamma_{\nu} \, .
\end{eqnarray}
It is apparent that the value of the above expression is not zero
nonperturbatively due to the non-Abelian property\cite{og2}. Then we
make clear the meaning of Eq.~(\ref{zong-result}) and also prove
that the claim made in Ref.~\cite{Gogohia} is incorrect.

Now an obvious question arises: whether there exists a truncation
for the quark-gluon vertex to guarantee the the validity of the
Slavnov-Taylor identity omitting the ghost effect? Recalling the
above discussion, we can infer that such a truncation does exist. To
make it realistic, we modify the ladder approximation of the
quark-gluon vertex equation as
\begin{eqnarray}\label{modi-ladder}
\Gamma_{\sigma}(p)& = & i\gamma_{\sigma} -
\left(C_{F}-\frac{C_{A}}{2} \right)g^{2}
\int\frac{d^{4}q}{(2\pi)^{4}}D_{\mu\nu}(k)\gamma_{\mu}S(q)
\Gamma_{\sigma}(q)S(q)\gamma_{\nu} \nonumber \\  & & \quad \; \;
+\frac{C_{A}}{2}g^{2}\int\frac{d^{4}q}{(2\pi)^{4}}D_{\mu\alpha}(k)
\Gamma_{\alpha\beta\sigma}^{3g}(k)D_{\beta\nu}(k)
\gamma_{\mu}S(q)\gamma_{\nu}\, .
\end{eqnarray}
It is apparent that the second term in the right-hand side of the
above equation is just the usual Abelian ladder approximation as
shown in Eq.~(\ref{ladddse}). An additional term is added in this
expression which is related to the nonperturbative three-gluon
vertex function. In this sense, we should prove that the
Slavnov-Taylor identity omitting the ghost effect is valid for
Eq.~(\ref{modi-ladder}). To reach such a point, we compare
Eqs.~(\ref{raindse}) and (\ref{modi-ladder}). It is evident that
Eq.~(\ref{raindse}) can be expressed in an equivalent form as
\begin{equation} \label{raindse-2}
S(p)^{-1}=i\gamma\cdot p + m_{0} + g^{2}(C_{F}-\frac{C_{A}}{2})
\Sigma(p)+g^{2}\frac{C_{A}}{2}\Sigma(p) \, ,
\end{equation}
where $\Sigma(p)$ is the quark self-energy in
Eq.~(\ref{self-energy}). Differentiating both sides of the above
equation with respect to external momentum, one can get the general
differential form of the quark equation
\begin{eqnarray}\label{diffdse-2}
\frac{\partial S(p)^{-1}}{\partial p_{\sigma}} & = &
i\gamma_{\sigma} -  \left(C_{F}-\frac{C_{A}}{2}\right
)g^{2}\int\frac{d^{4}q}{(2\pi)^{4}}D_{\mu\nu}(k)\gamma_{\mu}
S(q)\frac{\partial S(q)^{-1}}{\partial q_{\sigma}}S(q)\gamma_{\nu} \nonumber \\
& & \quad \, \,
+\frac{C_{A}}{2}g^{2}\int\frac{d^{4}q}{(2\pi)^{4}}D_{\mu\alpha}(k)
\Gamma_{\alpha\beta\sigma}^{3g}(k)D_{\beta\nu}(k)
\gamma_{\mu}S(q)\gamma_{\nu}\, ,
\end{eqnarray}
where the Slavnov-Taylor identity in Eq.~(\ref{relation-2}) has been
implemented for the third term in the right-hand side. Comparing
Eq.~(\ref{diffdse-2}) and Eq.~(\ref{modi-ladder}), one can obtain
the relation in Eq.~(\ref{WardI}). It shows apparently that the
Slavnov-Taylor identity for the quark-gluon vertex omitting the
ghost effects is valid for the modified ladder approximation with
the modified vertex function in Eq.~(\ref{modi-ladder}).

In summary, based on the Slavnov-Taylor identity for three-gluon
vertex, we have proved that the claim made by Gogohia\cite{Gogohia}
is incorrect in this letter. Then the claim can not be used to say
any thing about the internally inconsistent of the ladder
approximation to QCD. Comparing with the argument provided in
Ref.~\cite{zong}, our present statement is more physically and the
reason for the contradiction between the Wark-Takahashi identity
omitting ghost effects and the ladder approximation is represented
more clearly and generally. Moreover, we provide a modification for
the ladder approximation of the quark-gluon vertex which contains
the usual Abelian term and an additional non-Abelian term. With such
a modified ladder approximation to quark-gluon vertex and the
rainbow approximation to quark equation we prove that the
Slavnov-Taylor identity is valid for the quark-gluon vertex omitting
the ghost effects.

\bigskip

\bigskip


This work was supported by the National Natural Science Foundation
of China (NSFC) under contract Nos. 10425521 and 10575004, the
Major State Basic Research Development Program under contract No.
G2000077400, the research foundation of the Ministry of Education,
China (MOEC), under contact No. 305001 and the Research Fund for
the Doctoral Program of Higher Education of China under grant No
20040001010. One of the authors (YXL) thanks also the support of
the Foundation for University Key Teacher by the MOEC.


\begin{thebibliography}{99}
\bibitem{fnEuclidean} We use a Euclidean metric, with:
$\{\gamma_\mu,\gamma_\nu\} = 2\delta_{\mu\nu}$;
$\gamma_\mu^\dagger = \gamma_\mu$; $a \cdot b = \sum_{i=1}^4 a_i b_i$;
and ${\rm tr}[\gamma_5\gamma_\mu\gamma_\nu\gamma_\rho\gamma_\sigma]=
-4\,\epsilon_{\mu\nu\rho\sigma}\,, \epsilon_{1234}= 1\,.$  For a
spacelike vector $k_\mu$, $k^2>0$. Landau gauge is used in our
statements.
\end{thebibliography}
\end{document}